\documentclass[twocolumn,prl,amsmath,amssymb,showpacs]{revtex4}

\usepackage{graphicx}
\usepackage{dcolumn}
\usepackage{bm}
\usepackage{latexsym}
\usepackage{amstext}
\usepackage{amsxtra}
\usepackage{verbatim}
\usepackage[latin9]{inputenc}
\usepackage[OT4]{fontenc}
\usepackage[english]{babel}
\usepackage[usenames]{color}
\usepackage{epstopdf}
\usepackage{ifpdf}

\newcommand{\ER}{E_{\text{R}}}
\newcommand{\tn}{\textnormal}
\newcommand{\acrit}{a_{\tn{crit}}}
\newcommand{\add}{a_{\tn{dd}}}
\newcommand{\aho}{a_{\tn{ho}}}
\newcommand{\nbec}{N_\tn{BEC}}
\newcommand{\Nat}{N_\tn{at}}

\newcommand{\dlat}{d_\tn{lat}}

\newcommand{\be}{\begin{equation}}
\newcommand{\ee}{\end{equation}}

\newcommand{\rb}{{\bf r}}

\newcommand{\kb}{{\bf k}}

\begin{document}
\title{Dipolar Stabilization of an Attractive Bose Gas in a One Dimensional Lattice}
\author{S. Müller$^{1}$, J. Billy$^{1}$, E. A. L. Henn$^{1}$, H. Kadau$^{1}$, A. Griesmaier$^{1}$, M. Jona-Lasinio$^{2}$, L. Santos$^{2}$, and T. Pfau$^{1}$}
 \affiliation{
 $^1$ 5.Physikalisches Institut, Universität Stuttgart, Pfaffenwaldring 57, 70569 Stuttgart, Germany\\
 $^2$ Institut für Theoretische Physik, Leibniz Universität Hannover, 30167 Hannover, Germany
 }
 \date{\today}

\begin{abstract}
We experimentally show that dipolar interaction can stabilize otherwise unstable many-body systems like an attractive Bose gas. In a one dimensional lattice the repulsive dipolar on-site interaction balances negative scattering lengths up to $-17$ Bohr radii and stabilizes the $^{52}$Cr Bose-Einstein condensate. For reduced lattice depths, the dipolar stabilization turns into destabilization. We probe the full cross-over between the two regimes and our results are in excellent agreement with theoretical calculations, which reveal significant dipolar inter-site interactions. 
\end{abstract}


 \maketitle


Recent years have seen an increasing interest in the design and realization of novel quantum phases. In this regard, cold quantum gases play a central role~\cite{BlochReview,LewensteinReview} through the high level of experimental control that they allow. Internal as well as external degrees of freedom are controlled independently and one is able to tune the two-body short-range interaction that drives the main properties of such cold ensembles almost at will~\cite{FRReview}.

The toolbox for the design of possible novel quantum phases has been recently widened by a new ingredient: the dipolar interaction (DI). In strong contrast with the contact interaction, it has two assets, namely the anisotropy and the long-range character. In the last years, the DI in cold gases has been object of numerous experimental and theoretical investigations~\cite{BaranovReview,Lah2009}. Within this framework, quantum gases with strong dipolar interactions have been experimentally realized in systems of chromium Bose-Einstein condensates (BECs)~\cite{Lah2007,Koch2008,Beaufils2008}, ultra-cold heteronuclear molecules~\cite{Jila} and BECs with Rydberg excitations~\cite{Heidemann2008}, while weak dipolar effects have also been observed in alkali samples~\cite{Spinor2008,Fattori2008,Pollack2009}. On the theoretical side, unique self-organized structures have been predicted, such as vortex lattices of different symmetries~\cite{VortexLattice}, density modulated ground-states~\cite{Ronen2007, Dutta2007}, and supersolid phases in optical lattices~\cite{Goral2002,Danshita2009,Buehler2010}.

When confined in a quasi-two-dimensional (2D) geometry, the underlying mechanism responsible for self-organization in dipolar quantum gases is closely related to their roton-maxon excitation spectrum~\cite{Santos2003}. Furthermore, in a stack of such quasi-2D systems, long-range inter-site interactions are responsible for an enhancement of dipolar effects~\cite{Klawunn2009,Wang2008} and the amplification of the self-organized structures~\cite{Koeberle2009}. On the level of many-body physics, the dipolar atomic gases already fulfill the requirements of degeneracy and strong dipolar interactions. However, in these systems the structures described in Refs.~\cite{VortexLattice,Ronen2007,Dutta2007,Goral2002,Danshita2009,Buehler2010} are expected to form in a so far unreached range of parameters, where repulsive dipolar interaction counterbalances attractive contact interaction. Therefore investigating the stability of multi-site systems of highly oblate dipolar condensates is a crucial step towards the realization of novel quantum phases.
 
Experimentally, a one dimensional (1D) optical lattice oriented along the polarization direction of the dipoles allows for the realization of the desired oblate BECs with mainly repulsive DI. One expects the clouds in the individual lattice sites to be stable even at negative scattering lengths due to a strong dipolar stabilization~\cite{Koch2008}. But, as several lattice sites are populated, one also expects the mainly attractive inter-site interactions due to the long-range dipolar potential to destabilize the system. Because of this interplay between on-site stabilization and inter-site destabilization, it is \textit{a priori} unclear whether one can reach the regime of attractive contact interaction where different new quantum phases have been predicted to appear. 

In this work, we show that in a 1D optical lattice the strong dipolar repulsive on-site interaction can, in fact, stabilize a BEC of $^{52}$Cr atoms with attractive short-range interaction. We identify the stability threshold of the condensate for various lattice depths by measuring the BEC atom number when decreasing the \textit{s-}wave scattering length via magnetic Feshbach resonance. From small to large lattice depths we observe a continuous cross-over from a dipolar destabilized to a dipolar stabilized regime where we find a stable BEC even at negative scattering length. Indeed for deep lattices one can consider our system as a stack of mesoscopic quasi-2D BECs containing up to 2000 atoms in the central wells, strongly stabilized by the on-site dipolar interaction. Our measurements are in good agreement with numerical mean-field calculations, which include significant contribution of inter-site interactions mediated through the long-range dipolar potential.

 \begin{figure}[t]
\centerline{\includegraphics[width=0.8\linewidth]{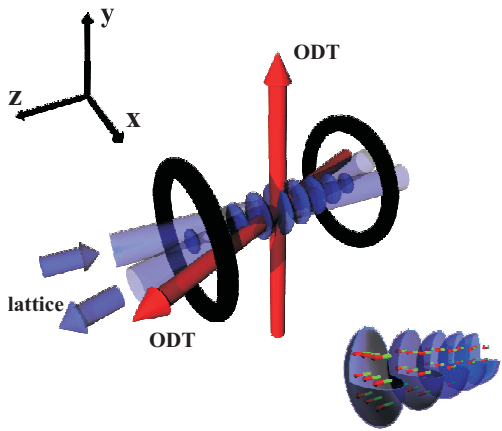}} \caption{Experimental setup. The measurements are performed in a 1D optical lattice (blue) with underlying crossed optical dipole trap (red). The magnetic field used to reach the Feshbach resonance is produced by Helmholtz coils (black) and polarizes the dipoles along the lattice direction ($z$). For deep lattices we obtain a stack of oblate dipolar BECs as depicted on the lower right.}
\label{fig:fig1ExpSetup}
\end{figure}

	Our experimental procedure is as follows: we produce a $^{52}$Cr BEC in a crossed optical dipole trap (ODT, $\nu_{x,y,z}=\left(440,330,290\right)\tn{Hz}$) at a magnetic field $B\simeq600\,\tn{G}$, where the scattering length is large and positive ($a\simeq90\,a_0$ with $a_0$ the Bohr radius). Dipoles are aligned along the strong magnetic field in $z$-direction. By changing the magnetic field strength $B$ in the vicinity of a Feshbach resonance (FR), we tune the \textit{s-}wave scattering length $a$ according to 
\begin{equation}
	a\left(\scriptsize{B}\right)=a_\text{bg}\cdot\left(1-\frac{\Delta}{B-B_0}\right),
	\label{eq:ScattLength}
\end{equation}
where $a_\text{bg}\simeq100 a_0$ is the background scattering length, $\Delta\simeq1.6\,\tn{G}$ the width and $B_0=589.1\,\tn{G}$ the center of the FR. After reducing the scattering length to $a=60\,a_0$, we load the BEC into the 1D optical lattice which is oriented along the polarization direction $z$. The lattice is produced by a $\lambda=1064\,\tn{nm}$ fiber laser in a nearly back-reflected geometry (full crossing angle $\alpha=10°$) with a lattice spacing $\dlat= 534\,\tn{nm}$, as illustrated in Fig.~\ref{fig:fig1ExpSetup}. The radial trapping frequencies ($\nu_{x,y}$) are kept constant during the lattice ramp by adjusting the power in the ODT beam in the $z$-direction. We then decrease the scattering length in $6\,\tn{ms}$ to reach its final value, where we hold the atoms for $t_{\tn{hold}}=2\,\tn{ms}$. Finally we switch off the optical trapping potential for a $6\,\tn{ms}$ time-of-flight (TOF) before taking an absorption image.

\begin{figure}[h]
\centerline{\includegraphics[width=\linewidth]{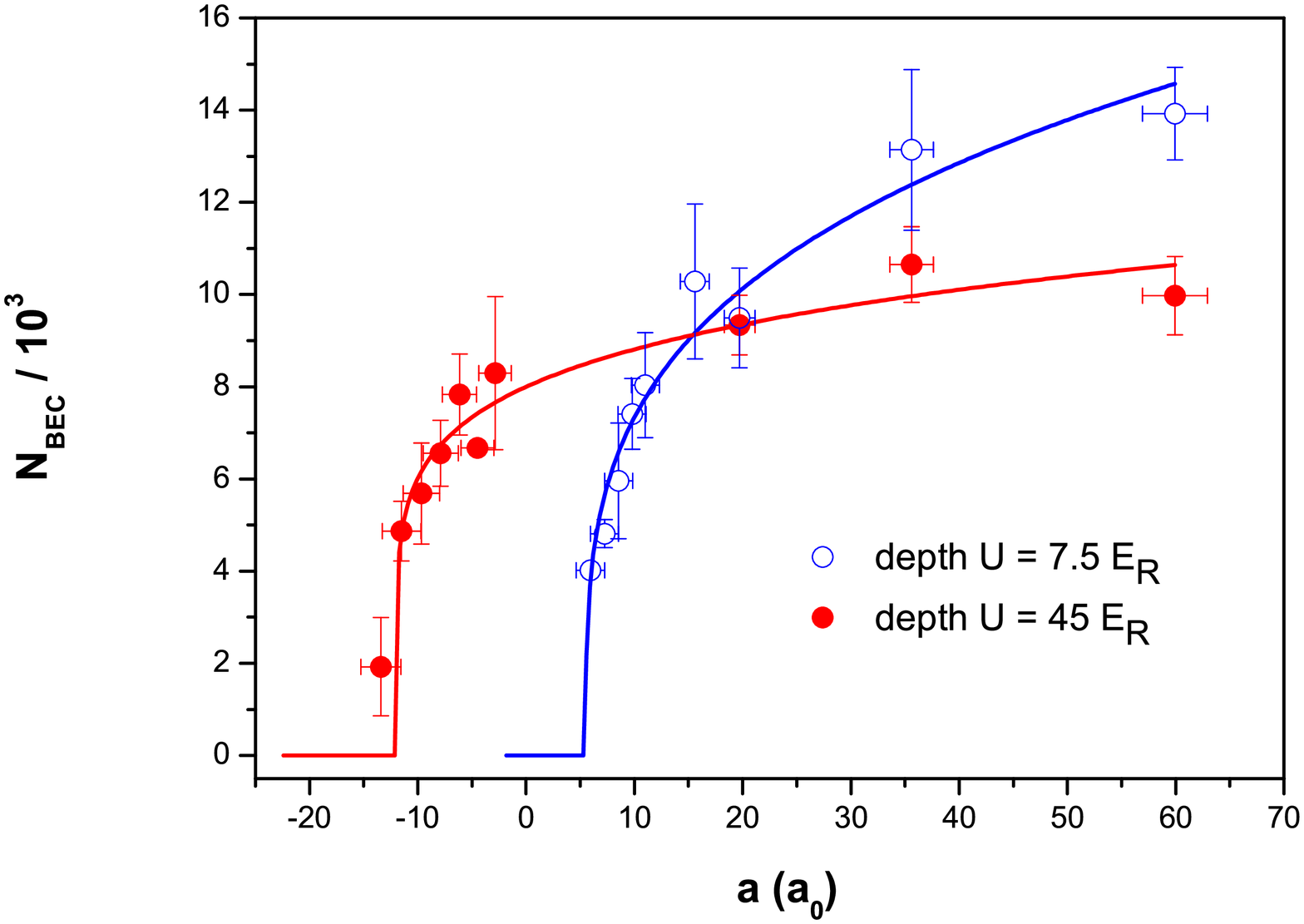}} \caption{Atom number versus scattering length in shallow and deep lattice. In a rather shallow lattice of depth $U=(7.5\pm0.8)\,\ER$ (open blue dots), the condensate becomes unstable at $\acrit=(6.5\pm1.9)a_0$, while in a deep lattice at $U=(45\pm5)\,\ER$ (filled red dots) we even find a stable BEC until $a=(-13.2\pm2.5)\,a_0$. The solid lines are fits to the data using the arbitrarily chosen form $\nbec = \tn{max}\left\{0,N_0\!\cdot\!\left(a-\acrit\right)^\beta\right\}$, from which we extract the critical scattering length $\acrit$ (typically $\beta\simeq 0.2$).}
\label{fig:fig2StabilityMeasurements}
\end{figure}

To extract the BEC atom number $\nbec$ after TOF, the recorded 2D density distribution is integrated along the \textit{z}-direction and we perform a 1D bimodal fit. When the final scattering length is much larger than the critical value $\acrit$, we typically measure $\nbec\simeq 15.000$ ($\nbec\simeq 10.000$) in a shallow (deep) lattice, while the atom number before loading the lattice is always $\Nat\simeq 20.000$.  Getting close to the instability point, we observe a fast decrease in the atom number as it is shown in Fig.~\ref{fig:fig2StabilityMeasurements} for two different values of the lattice depth. We finally extract the critical scattering length from an empirically chosen function as described in Ref.~\cite{Koch2008}. Although atom losses are enhanced in a deep lattice due to the larger mean trapping frequency, they do not affect the determination of the stability threshold.

\begin{figure}[h]
\centerline{\includegraphics[width=\linewidth]{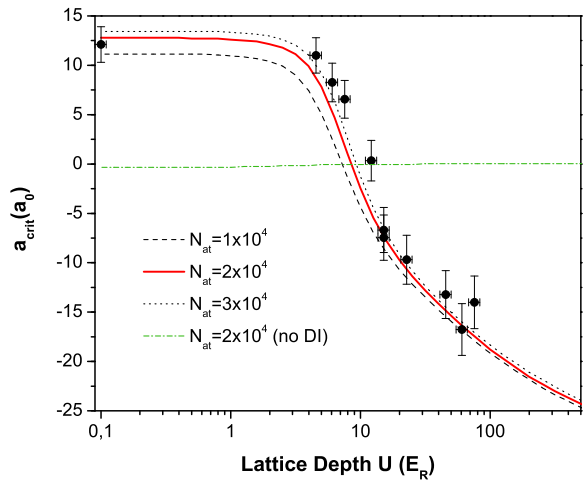}} \caption{Stability diagram of the dipolar condensate in the 1D optical lattice. The critical scattering length $\acrit(a_0)$ is plotted versus the lattice depth $U (\ER)$. The lines are results of the numerical simulations for different atom numbers. The full cross-over from a dipolar destabilized ($\acrit>0$) to a dipolar stabilized ($\acrit<0$) regime is observed. At $U\approx 10\,\ER$ a purely dipolar interacting BEC ($\acrit=0$) can be stabilized. The dash dotted line (green) shows the simulated critical scattering length disregarding dipolar interaction.}
\label{fig:fig3StabilityDiagram}
\end{figure}

Figure \ref{fig:fig3StabilityDiagram} shows the stability diagram of a dipolar $^{52}$Cr BEC in a 1D optical lattice. The critical scattering length $\acrit$ is measured for different lattice depths in the range $U=0-80\,\ER$ (recoil energy, $\ER=\hbar^2\pi^2/(2m\dlat^2)$, with $m$ the atomic mass). We find positive $\acrit$ until $U\simeq 10\,\ER$ and down to $\acrit=(-17\pm3)a_0$ in a deep lattice potential. The datapoints are in very good agreement with numerical simulations based on the non-local non-linear Schrödinger equation
\begin{multline} 
i\hbar \frac{\partial}{\partial t} \Psi(\rb,t) =\left[-\frac{\hbar^2}{2m} \nabla^2
+ V_{\rm ext}(\rb)
+g\Nat|\Psi(\rb,t)|^2 \right. \\ \left.
+\Nat\int d\rb^\prime\,V_{\rm dd}(\rb-\rb^\prime)|\Psi(\rb^\prime,t)|^2\right]
\Psi(\rb,t), 
\label{eq:schrod1}
\end{multline} 
where $g = 4\pi\hbar^2 a(B)/m$. The potential $V_{\rm ext}(\rb)=  U \sin^2(\pi z/\dlat) + m \sum_{i=x,y,z}4\pi^{2}\nu_i^2 r_i^2/2$ results from the 1D optical lattice and
the 3D harmonic confinement given by the ODT. The DI potential is given by 
$V_{\rm dd}(\rb) = \frac{\mu_0 \mu^2}{4\pi} \frac{1-3(\hat r \cdot \hat z)^2}{r^3}$ 
(with $\hat r= \rb /r$), where $\mu_0$ is the vacuum permeability and $\mu$ the magnetic dipole moment ($\mu=6\mu_\tn{B}$ for $^{52}$Cr with $\mu_\tn{B}$ the Bohr magneton).

The DI term is calculated by using the convolution theorem and the Fourier transform 
of the DI potential, $\widetilde V_{\rm dd}(\kb) = \frac{2\mu_0 \mu^2}{3} \left(1- \frac{3}{2}|\hat k \times \hat z|^2\right)$~\cite{Lah2009}. We evaluate the ground-state of the system using a full 3D simulation of Eq.~(\ref{eq:schrod1}) in imaginary time. The full 3D character of the calculations allows us to determine the critical scattering length $\acrit$ for all lattice depths, ranging from $U=0$ to very deep lattices. 
Alternatively, we evaluate Eq.~(\ref{eq:schrod1}) in real time, and simulate the actual experimental 
sequence described above. The time-dependent results are basically undistinguishable from those obtained from ground-state considerations in imaginary time. This implies that the experimental results recover the stability threshold of a dipolar BEC even for deep lattices, for which reaching experimentally the ground state becomes questionable. Furthermore, the comparison of simulations for different atom numbers (Fig.~\ref{fig:fig3StabilityDiagram}) shows that the stability threshold depends only weakly on the initial BEC atom number $N_\tn{at}$, in particular in the deep lattice regime.

In order to understand the results shown in Fig. \ref{fig:fig3StabilityDiagram}, we recall that the stability of a dipolar BEC strongly depends on its geometry~\cite{Koch2008}. It is also convenient to introduce a length scale associated to the DI, $\add=m\mu_0\mu^2/12\pi\hbar^2$ ($\simeq 15\,a_0$ for $^{52}$Cr). When the lattice is absent, the ODT determines the shape of the BEC and therefore its critical scattering length. In our case, the trapping frequencies are chosen such that both trap aspect ratios $\lambda_{x,y}=\frac{\nu_z}{\nu_{x,y}}$ are smaller than unity. In this essentially prolate trap, the dipolar interaction strongly destabilizes the condensate. The BEC becomes then unstable at a positive scattering length $\acrit=(12\pm 2)\,a_0$, close to the value $a=\add$, which is the expected critical scattering length in a prolate trap in the limit $\Nat\add / \aho\gg 1$~\cite{Koch2008}, with $\aho$ the mean harmonic oscillator length.

On the contrary, for a deep lattice, the relevant aspect ratio is that of the single site trap, which becomes largely oblate. Hence, for a growing lattice depth, we observe a smooth cross-over from a dipolar destabilized ($\acrit>0$) to a dipolar stabilized ($\acrit<0$) regime,  characterized in Fig. \ref{fig:fig3StabilityDiagram} by a rapid decrease of $\acrit$ for depths $U\sim5-10\,\ER$. For larger depths, at which inter-site hopping is negligible, the decrease of $\acrit$ becomes much slower and we find a minimum critical scattering length of $\acrit=(-17\pm 3)\,a_0$ at $U=60\,\ER$. This is in strong contrast with the behavior of a purely contact interacting BEC (with same parameters otherwise), for which simulations (see Fig.~\ref{fig:fig3StabilityDiagram}) show only weak dependence of $\acrit$ on the lattice depth, with $\acrit\simeq 0\,a_0$ in the full range~\cite{ComparisonAlkali}. In this respect, we show here that the DI stabilizes an attractive Bose gas.

In the deep lattice region ($U\geq15\,\ER$), a naive picture based on the stability criterion for a single oblate condensate would suggest that $\acrit$ should rapidly approach $-2\add$. Experimentally, we observe a stable condensate until $\acrit\sim -17\,a_0$ only. Indeed in a deep lattice, our system becomes a stack of quasi-2D pancake-like condensates with inter-site interactions mediated by the long-range DI. Due to the anisotropy of the DI, the inter-site interactions have a destabilizing character, and as a result $\acrit$ is considerably higher than the expected $-2\add$. Note that this inter-site destabilization requires an inhomegeneous BEC, since two parallel infinite homogeneous quasi-2D condensates would present zero-averaged inter-site DI~\cite{Pikovski2010}.


\begin{figure}[h]
\centerline{\includegraphics[width=.8\linewidth]{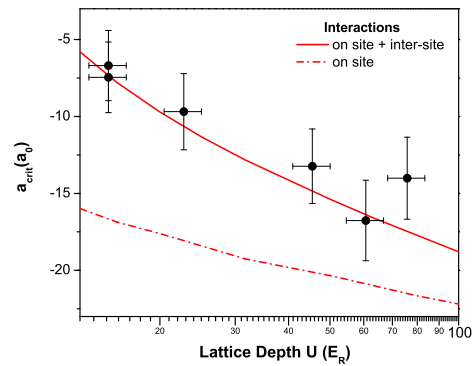}} \caption{Inter-site coupling mediated by dipolar interactions: zoom on Fig.~\ref{fig:fig3StabilityDiagram} in the deep lattice regime. Inter-site hopping is negligible on experimental time scales. Solid line: full 3D numerical simulation. Dashed line: simulation with the truncated dipolar potential (see text), for which inter-site coupling is not taken into account. The deviation of the simulation with the truncated potential to the experimental data indicates significant mean-field energy contribution from long-range inter-site interactions.}
\label{fig:fig4ZoomDeepLattice}
\end{figure}


In the following, we further investigate the destabilizing role of the dipolar inter-site interactions. In this respect, strong insight is gained by introducing a truncated DI potential, $V_{\rm dd}^{\rm box}(\rb) = V_{\rm dd}(\rb)\left[\Theta\left((\rb\cdot\hat z) + d_{lat}/2\right) - \Theta\left((\rb\cdot\hat z) - d_{lat}/2\right)\right]$, where $\Theta(\xi)$ is the Heaviside function. Such cut-off is implemented 
following a similar procedure as that of Ref.~\cite{Bortolotti2006}. For deep lattices the cut-off in $V_{\rm dd}^{\rm box}(\rb)$ effectively amounts to remove inter-site DI, while still taking into account the short-range and long-range on-site interactions. In Fig.~\ref{fig:fig4ZoomDeepLattice} we compare the stability threshold calculated with $V_{\rm dd}^{\rm box}(\rb)$ to the data and the full simulations in the deep lattice regime ($U\geq15\,\ER$). The calculations with the truncated dipolar potential show a substantial deviation to the experimental data~\cite{Limit}: for instance, at $U\simeq 20\,\ER$ the difference is $\Delta\acrit\simeq 8\,a_0$, which is more than three times our standard deviation. This discrepancy, in addition with the very good agreement of the full simulations with the measurements, provides a clear signature of the presence of inter-site interactions. Note that these inter-site interactions reduce the stabilization effect of the on-site repulsive DI~(e.g. for $U=20\,\ER$, $|\acrit|$ is reduced by close to $50\%$), indicating that they contribute significantly to the overall energy of the dipolar BEC in our experiments. 

In conclusion we have experimentally shown the stabilization of a polarized dipolar Bose gas with attractive contact interaction. We produced the required large trap aspect ratio by loading the $^{52}$Cr BEC into a one dimensional optical lattice potential. The measurements are well described by numerical mean-field calculations using the non-local non-linear Schrödinger equation. In the regime of negligible inter-site hopping, the data supports significant long-range inter-site interactions.

The successful stabilization of a dipolar quantum gas with attractive short-range interactions paves the way to the investigation of the roton-maxon excitation spectrum expected for such systems~\cite{Santos2003, Klawunn2009, Nath2010} and the related self-organized structures. Furthermore, the realization of a stack of stable mesoscopic ensembles interacting through long-range dipolar interactions, together with the possibility to tailor multi-well potentials at will~\cite{Henderson2009,Zimmermann2011}, is particularly promising for the experimental realization of new quantum phases relying on the interplay between on-site and long-range inter-site interactions, as predicted in Ref.~\cite{Lah2010}.

We thank K. Rz\k{a}\.{z}ewski, K. Paw\l owski and D. Peter for fruitful discussions and P. Weinmann and T. Maier for general contributions. The Stuttgart group is supported by the German Research Foundation (DFG, through SFB/TRR21), the German-Israeli Foundation and contract research `Internationale Spitzenforschung II' of the Baden-Württemberg Stiftung. The Hannover group acknowledges financial support by the DFG and the Cluster of Excellence QUEST. E.A.L.H. acknowledges support by the Alexander von Humboldt-Foundation.



\begin{thebibliography}{99}
\bibitem{Limit}
The stability thresholds calculated with $V_{\rm dd}(\rb)$ and $V_{\rm dd}^{\rm box}(\rb)$ reach their common asymptotic value $\acrit\simeq -2\add$ only slowly, since the on-site trap aspect ratio is limited and depends on the lattice depth as $\left(U/\ER\right)^{1/2}$.
\end{thebibliography}
\end{document}